\input epsf
\input harvmac.tex

\def\capt#1{\narrower{
\baselineskip=14pt plus 1pt minus 1pt #1}}

\lref\Baxter{Baxter, R.J.: Exactly solved models
in Statistical Mechanics. London: Academic Press, 1982}

\lref\lusher{Luther, A. and Peschel, I.: Calculation of critical exponents
in two dimensions from quantum field theory in one dimensions.
Phys. Rev. {\bf B12}, 3908-3917 (1975)}

\lref\Kadan{Kadanoff, L.P. and Brown, A.C.: Correlation functions on
the critical lines of the Baxter and Ashkin-Teller models.
Ann. Phys. (N.Y.)
{\bf 121}, 318-342 (1979)}

\lref\osh{Oshikawa, M. and  Affleck I.: Theory of the field-induced gap
in $S=1/2$ antiferromagnetic chain. Preprint (1997), cond-mat/970685}

\lref\tse{Essler, F.H.L. and Tsvelik, A.M.: Dynamical magnetic
susceptibilities in $Cu$ Benzoate.
Preprint (1997),  cond-mat/9708208 }

\lref\AGSZ{Affleck, I., Gepner, D., Schultz H.J. and
Ziman, T.: Critical behavior of spin $s$ Heisenberg 
antiferromagnetic chains:
analytic and numerical results.
J.Phys. {\bf A22}, 511-529  (1989)}

\lref\Kor{Korepin, V.E., Bogoliubov, N.M. and Izergin, A.G.:
Quantum inverse  scattering method and correlation functions.
Cambridge University Press, 1993}

\lref\Korr{Essler, F.H.L., Frahm, H., Izergin, A.G. and
Korepin V.E.: Determinant representation for correlation functions of
spin-${1\over 2}$ XXX and XXZ Heisenberg magnets. Comm. Math. Phys.
{\bf 174}, 191-214 (1995)}

\lref\Korra{Korepin V.E,  Izergin, A.G., Essler, F.H.L. and Uglov, D.B.:
Correlation function of the 
spin-${1\over 2}$ XXX antiferromagnet.
Phys. Lett. {\bf A190}, 182-184 (1994) }

\lref\afff{Affleck, I: Field theory methods and
quantum critical phenomena. In:
Br${\acute{\rm e}}$zin, E. and Zinn-Justin J. (eds)
Fields, Strings, Critical Phenomena. Proceedings, 563-640.
North-Holland,
Amsterdam, 1990}

\lref\MiW{Jimbo, M. and Miwa, T.: Quantum KZ equation
with\ $|q|=1$\ and correlation functions
of  the XXZ model in the gapless regime.
J. Phys. {\bf A29}, 2923-2958 (1996)}

\lref\Hasenf{Hasenfratz, P.,  Maggiore, M. and Niedermayer, F.:
The exact mass gap of the $O(3)$ and $O(4)$ non-linear $\sigma$-models
in $d=2$.  
Phys. Lett. {\bf 245}, 522-528 (1990)\semi
The exact mass gap of the $O(N)$ $\sigma$-model
for arbitrary $N\geq 3$  in $d=2$. Phys. Lett. {\bf 245}, 529-532 (1990)}

\lref\yangaa{Yang, C.N. and Yang C.P.:
One dimensional chain of anisotropic spin-spin interaction.
I. Proof of Bethe's hypothesis for ground state in a finite system.
{\bf 150},
321-327 (1966)\semi
II. Properties of the ground state energy per lattice site in a finite
system. {\bf 150},
327-339 (1966)}

\lref\zin{ Itzykson, C. and Drouffe, J.-M.: Statistical field theory.
Cambridge University Press, 1989}

\lref\sosage{Fateev, V.A, Onofri, E. and Zamolodchikov Al.B.:
The Sausage model (integrable deformations of $O(3)$ Sigma-model).
Nucl. Phys. {\bf B406}, 521-565 (1993) }

\lref\Af{Eggert, S., Affleck, I. and Takahashi, M.:
Susceptibility of the spin ${1\over 2}$ Heisenberg antiferromagnetic
chain. Phys. Rev. Lett. {\bf 73}, 332 (1994)}

\lref\LZ{Lukyanov, S. and Zamolodchikov, A.:
Exact expectation values of local fields
in quantum sine-Gordon model.
Nucl. Phys. {\bf B493}, 571-587 (1997)}

\lref\affleck{Affleck, I.: unpublished }

\lref\BLZ{Bazhanov, V.V., Lukyanov, S.L. and Zamolodchikov A.B.:
Integrable structure of Conformal Field Theory,
quantum KdV theory and Thermodynamic Bethe Ansatz.
Comm. Math. Phys. {\bf 177}, 381-398 (1996)}

\lref\BLZZ{Bazhanov, V.V., Lukyanov, S.L. and Zamolodchikov A.B.: 
Integrable structure of Conformal Field Theory,
II. Q-operator and DDV equation.
Comm. Math. Phys. {\bf 190} 247-278 (1997)}

\lref\afa{Affleck, I.: Critical behavior of two-dimensional systems with
continuous symmetries. Phys. Rev. Lett. {\bf 55}, 1355-1358 (1985)}  

\lref\hallberg{Hallberg, K.A., Horsch, P.
and Mart${\acute {\rm i }}$nez, G.:
Numerical renormalization-group study of the correlation functions of the 
antiferromagnetic spin-${1\over 2}$ Heisenberg chain.
Phys. Rev. {\bf B52}(2),  R719-722 (1995) } 

\lref\lin{Lin, H.Q. and Campbell, D.K.: Spin-spin correlations in
the one-dimensional spin-${1\over 2}$, antiferromagnetic Heisenberg chain.
J. Appl. Phys. {\bf 69}, 5947-5949 (1991)} 

\lref\Zarn{Zamolodchikov, Al.B.: Mass scale
in the Sine-Gordon model and its
reductions. Int. J. Mod. Phys. {\bf A10}, 1125-1150  (1995) }

\lref\zamol{Zamolodchikov, Al.B.: unpublished }

\lref\barber{Aclaraz, F.C., Barber, M.N. and Batchelor T.M.:
Conformal invariance and the spectrum of the XXZ chain.
Phys. Rev. Lett. {\bf 58}, 771-774 (1987)}

\Title{\vbox{\baselineskip12pt\hbox{RU-97-103}
                \hbox{cond-mat/9712314}}}
{\vbox{\centerline{}
\centerline{Low energy effective Hamiltonian }
\centerline{}
\centerline{  for   the XXZ spin chain }}}

\centerline{}
\centerline{ Sergei Lukyanov}
\centerline{}
\centerline{Department of Physics and Astronomy,
Rutgers University}
\centerline{ Piscataway,
NJ 08855-0849, USA}
\centerline{and}
\centerline{L.D. Landau Institute for Theoretical Physics}
\centerline{Kosygina 2, Moscow, Russia}
\centerline{}

\centerline{}

\centerline{\bf{Abstract}}

Coupling constants for the most relevant  terms in the low energy 
effective Hamiltonian of the XXZ spin chain are derived.
Using this result  we study the 
low energy (low temperature, weak magnetic field)
thermodynamics, finite size effects and subleading long distance
asymptotics of   correlation functions.

\Date{December, 97}

\eject

\newsec{ Introduction}

An effective Hamiltonian approach is an important
tool for a  qualitative understanding
of low energy effects in Quantum Field Theory 
and Statistical Mechanics\ \zin.
The quantitative description involves  some problems.
First, it is not generally   clear how to derive coupling constants of  an
effective  Hamiltonian
in terms of  parameters of a 
microscopic Hamiltonian.
(The latter is in common use
for the  study  of a  high energy
behavior.) When it is not possible to relate the large scale physics
to the microscopic parameters directly, one  
uses instead  the  coupling constants of the  effective  Hamiltonian as  
phenomenological parameters of the  model under consideration.   
A related problem concerns a  control over an  accuracy of   perturbative 
expansions.
An  effective Hamiltonian density is a  series which
involves an infinite set
of local irrelevant fields and the corresponding perturbation theory is
unrenormalizable. Thus, it is impossible  to make a uniform (on the
energy scale)   
estimate  of an accuracy in  a  given order of the
perturbation theory generated by the  
low energy effective Hamiltonian.

Two dimensional exactly solvable models provide unique opportunity to
refine our understanding of the effective Hamiltonian approach.
Recently a  significant  progress was made in this field of research.
In the works\ \Hasenf,\ \sosage,\ \Zarn\ the method
have been  developed,
which enables one to find an  exact relation between
short and long distance asymptotic behaviors of physically
interesting quantities.
As a result
the coupling
constants of the  effective
Hamiltonians are expressed  explicitly
in terms of  the microscopic parameters for many
interesting integrable  models.

This work constitutes an  attempt to  apply the  effective 
Hamiltonian approach to a  quantitative study of low-energy effects in
the XXZ Heisenberg spin chain\ \Baxter,\ \Kor.
The model was chosen for several
reasons. First of all, the method of the papers\ \Hasenf,\ \sosage,\ 
\Zarn\ is suitable for  calculation  of
the XXZ effective Hamiltonian. 
Next,  due to more than sixty  years of  study,
a huge
amount of numerical data is  available. Finally, the 
XXZ Heisenberg spin chain still comes to the  attention
of  Condensed Matter physicists\ \ \osh,\ \tse.

In  Section 2, we fix our notations and present
the coupling constants of the most relevant
terms in the
effective XXZ Hamiltonian.
In Sections 3  and 4, relying on the 
effective Hamiltonian,  we develop  the perturbation theory
to study the
low energy thermodynamics and finite size corrections
to vacuum energies. In Section 5, we discuss the  subleading
long distance asymptotics of the  two-point correlation function of
the XXZ spin chain.

\newsec{Low energy effective Hamiltonian}
In this work we consider the XXZ spin  chain\ \Baxter,\ \Kor,
\eqn\xxz{{\bf H}_{XXZ}=-{J\over{2}}
\sum_{k=1}^{N}\big(\, 
\sigma_{k}^{x}
\sigma_{k+1}^{x}+\sigma_{k}^{y}\sigma_{k+1}^{y}+\Delta\,
(\sigma_{k}^{z}\sigma_{k+1}^{z}-1)\, \big)\ ,}
$\sigma_{k}^{x}, \sigma_{k}^{y}$ and $\sigma_{k}^{z}$
are the  Pauli matrices
associated with the site $k$
of the chain. We assume that $N$
is an even integer, and
supplement\ \xxz\ with the boundary conditions
\eqn\ksjdhgy{\eqalign{&\sigma^{\pm}_1=e^{\pm
2\pi i\theta}\, \sigma^{\pm}_{N+1}\ ,
\cr &\sigma^{z}_1=\sigma^{z}_{N+1}\ ,}}
with the  real parameter $0<\theta<1$.
Let the parameter $J$ be positive and 
\eqn\jsudt{-1<\Delta<1\  .}
The regime\ \jsudt\ is usually referred to  as the ``disorder regime''.
Notice that the unitary transformation, 
\eqn\jshy{{\bf U}^+\sigma^x_k\, {\bf U}=(-1)^k\sigma^x_k\, , \ \ \ 
{\bf U}^+\sigma^y_k\,  {\bf U}=(-1)^k\sigma^y_k\, , \ \ \ 
{\bf U}^+\sigma^z_k\, {\bf U}=\sigma^z_k\ ,}
can be equivalently described by the substitution
\eqn\hsydrtr{ J\to-J\ , \ \ \ \Delta\to -\Delta\ .}
In particular, the model\ \xxz\ 
with  $\Delta=-1$ is  unitary equivalent to the
$SU(2)$-invariant antiferromagnetic XXX spin chain. 
It will be  convenient for us  to parameterize the constants
in  the following manner
\eqn\hsgdta{\eqalign{&\Delta=\cos(\pi\beta^2)\ ,\cr
&J={{1-\beta^2}\over{ \,\sin(\pi\beta^2)}}\ a^{-1} \ ,}}
with $0<\beta^2\leq 1,\ a>0$. 

In the  thermodynamic limit
\eqn\jsdgre{
N\to \infty\, ,
\ \ \ \  a\to 0\, ,\ \ \ \  L=Na -{\rm fixed}\ \ \  \ \ (N-{\rm even})   }
the XXZ spin chain in the disorder regime\ \jsudt\ renormalizes
to the continuous quantum field
theory -- the Gaussian model\ \lusher,\  \Kadan,\ \barber, 
\eqn\sreew{\lim_{a\to 0}\, \Big(\, {\bf H}_{XXZ}-
a^{-1}\,  L\   {\cal E}_{0}\, \Big)=  {\bf H}_{Gauss}\ .}
Here\ \yangaa,
\eqn\jsdhfgr{{\cal E}_{0}= -{2\over \pi a}\ (1-\beta^2)\,  
\int_{0}^{\infty} d t\ 
{ {\rm sinh}\big(\beta^2 t  \big)\over
{\rm sinh}( t)\,
{\rm cosh}\big((1-\beta^2) t\big )}  \ .}
To define the Hamiltonian
of the Gaussian model let us introduce the set of
operators satisfying the commutation relations
\eqn\ksjdgt{\eqalign{& [a_{n}, a_{m}]=2 n \, \delta_{n+m,0}\, , \
\ \ \ \ [Q, P]=i\ ,\cr
&[{\bar a}_{n}, {\bar a}_{m}]=2 n \, \delta_{n+m,0}
\, , \  \ \ \ \ [{\bar Q}, {\bar P}]=i\ .}}
The Heisenberg algebra admits representation in the 
Fock spaces, i.e. the spaces generated by the action of
$a_{n}, {\bar a}_{k}$ with $n, k<0$  on the ``vacuum states''
$|\, p,{\bar p} \, \rangle $ which obey the equations
\eqn\hsgdreww{\eqalign{&a_n|\, p,{\bar p} \, \rangle\,
={\bar a}_{n} |\, p,{\bar p} \, \rangle\, =0\, 
\ \ \ {\rm for}\ \ \ \   n>0\ ,\cr
& P|\, p,{\bar p} \, \rangle=p\, |\, p,{\bar p} \, \rangle\, ,\ \ \ \  
{\bar P}|\, p,{\bar p} \, 
\rangle={\bar p}\, |\, p,{\bar p}\, \rangle \ .}} 
Then
\eqn\lsowija{{\bf H}_{Gauss}={2 \pi\over L} \Big\{\,
P^2+{\bar P}^2-{1\over 12} +{1\over 2}\, \sum_{n> 0} \big( a_{-n} a_n +
{\bar  a}_{-n} {\bar  a}_n\big)\ \Big\}\ .}
This operator acts in the Hilbert space
\eqn\jsudyt{{\cal H}=\oplus_{p, {\bar p}}\,  {\cal F}_{p,{\bar p}}\ ,}
with 
\eqn\hsgdre{\eqalign{&p-{\bar p}=s\, \beta\ ,\cr
& p+{\bar p}=(\theta+m)\, \beta^{-1}\ ,}}
here  $m$ and $s$  are  arbitrary integers.
It is convenient  to introduce the fields
\eqn\wrwt{\eqalign{&\phi(x)=Q+{4\pi x\over L}\, P-
i\sum_{n\not= 0} \, {a_n\over n} e^{{2\pi\over L} i x n}\ ,\cr
&{\bar \phi}(x)={\bar Q}-  {4\pi x\over L}\, {\bar P}-
i\sum_{n\not= 0} \, {{\bar a}_n\over n} e^{-{2\pi\over L} i x n}\ .}}
The Hamiltonian\ \lsowija\ can be written in the form
\eqn\hdgftr{{\bf H}_{Gauss}=\int_{0}^L {d x\over 2 \pi}\, 
\Big\{\, {\bf T}+{\bar {\bf T}}\, \Big\}\ ,}
where
\eqn\wreewt{\eqalign{&{\bf T}(x)={1\over 4} :(\partial_x
\phi)^2:-{\pi^2\over 6 L^2}\ ,\cr
&{\bf T}(x)={1\over 4} :(\partial_x
{\bar \phi})^2:-{\pi^2\over 6 L^2}\ . }}
To describe the bosonization  of the matrix $\sigma_{k}^{a}$,
we define the fields
\eqn\jdhfg{\eqalign{&\varphi(x)=\phi-{\bar \phi}\ \cr
&\tilde{\varphi}(x)=\phi+{\bar \phi}\ .}}
In the leading order in\ $a$\ \lusher\foot{In the
case of $\sigma_k^z$,\ Eq.(2.19) gives the leading order in $a$
for $0<\beta^2<{1\over 2}$.}, 
\eqn\shdttt{\eqalign{&\sigma_k^{\pm}\simeq a^{{\beta^2\over 2}}\ 
\sqrt{{F\over 2}}\
e^{\pm i{\beta\varphi\over 2} }(x)\ ,\cr
&\sigma_k^{z}\simeq {a\over 2\pi \beta}\ \partial_x \tilde{\varphi}(x)\  ,
}}
with $x=ka$.
To give a precise meaning to the constant 
$F$ one  needs to specify the normalization of the exponential fields.
We will normalize  exponential  operators
in accordance with the short distance expansion 
\eqn\ksjdy{\eqalign{
&e^{i\alpha\varphi}(x)\, e^{-i\alpha\varphi}(x')\ 
\cr
&e^{i\alpha\tilde{\varphi}}(x)\, e^{-i
\alpha\tilde{\varphi}}(x')\ }
\biggr\}\longrightarrow
|x-x'|^{-4 \alpha^2}\ \ \ \ {\rm as}\ \  |x-x'|\to 0\ ,}
for an arbitrary $\alpha$.
In this normalization the constant $F=F(\beta^2)$ was
found in Ref.\ \LZ
\eqn\xxzfactor{\eqalign{F=&{1\over 2\, (1-\beta^2)^2}\
\biggl[\, { \Gamma\big({\beta^2\over 2-2\beta^2}\big)\over
2\sqrt\pi\  \Gamma\big({1\over 2-2\beta^2}\big)}\,
\biggr]^{\beta^2}\times\cr &
{\rm exp}\biggl\{-
\int_{0}^{\infty} {d t\over t}\ \Big(\,
{ {\rm sinh}\big(\beta^2 t  \big)\over
{\rm sinh}( t)\,
{\rm cosh}\big((1-\beta^2) t\big )}-
\beta^2\,   e^{-2 t}\, \Big)\biggr\}\ . }}
The Hamiltonians\ \xxz\ and \hdgftr\ coincide 
in the lowest nontrivial order of the lattice parameter $a$\ \sreew.
In the next orders irrelevant fields contribute to the
density of ${\bf H}_{XXZ}$.
These irrelevant  fields should be local with respect to the 
spin fields\ \shdttt\ and commute with the operator
\eqn\usyt{{\bf S}^z={1\over 2}
\, \sum_{s=1}^N\sigma^z_s=(P-{\bar P})\, \beta^{-1}\ .}
A simple analysis  shows that the most important corrections have
the form\ \barber
\eqn\sew{\eqalign{
{\bf H}_{XXZ}=&  a^{-1} L\,  {\cal E}_0+ \int_{0}^L {d x\over 2 \pi}\, 
\Big\{\, {\bf T}+{\bar {\bf T}}-
a^2\, \Big( \lambda_{+}\,  {\bf T} {\bar {\bf T}}+ \lambda_{-}\,
\big( {\bf T}^2+{\bar {\bf T}}^2\big )\Big) +\cr &
\ \ \ \ \ \ \ \ \ \  \ \ \ \ \ \ \ a^{{2\over\beta^2}-2}\  \lambda\, 
\cos\big(\tilde{\varphi}/\beta\big)+...\, \Big\}\ .}}
Here, the field ${\bf T}^2 \ ({\bar {\bf T}}^2)$
is defined as a regular part of the
operator product expansion of two operators ${\bf T}\  ({\bar {\bf T}})$.
The dots in\ \sew\ and below  mean a contribution of higher
dimensional local counterterms.
In the Appendix we describe the  calculation of
the coupling constants in the low energy effective Hamiltonian\ \sew,
\eqn\jsdhgyt{\eqalign{
&\lambda={4\ \Gamma(\beta^{-2})\over
\Gamma(1-\beta^{-2})}\ \biggr[\,{ \Gamma\big(1+
{\beta^2\over 2-2\beta^2}\big)
\over 2\, \sqrt\pi \ \Gamma\big(1+{1\over 2-2\beta^2}\big)}\, \biggl]^{
{2\over \beta^{2}}-
2}\ ,\cr
&\lambda_+={1\over 2 \pi}\, \tan\big({\pi\over 2-2\beta^2}\big)\ ,\cr
&\lambda_-={\beta^2\over 12\pi}\ {\Gamma\big({3\over 2-2\beta^2}\big)\,
\Gamma^3\big({\beta^2\over 2-2\beta^2}\big) \over
\Gamma\big({3\beta^2\over 2-2\beta^2}\big)\,
\Gamma^3\big({1\over 2-2\beta^2}\big)}\  . }} 
For the  XXX spin chain ($\Delta=-1$), 
\eqn\gstdr{\eqalign{
&\lambda_+=0\ , \cr
&\lambda_-={\sqrt3\over  4\pi}\ .}}
We  discuss how to treat  ``$\lambda$-term'' from\ \sew\ at the
limit $\beta^2\to 1$
in the next section.

It is important to note that except for
the ``$\lambda_-$-term'',  the  Hamiltonian\ \sew\ 
coincides with low energy effective  Hamiltonian for
the sausage model with the value of the $\theta$-angle
equals to $\pi$ \sosage.\foot{V.A. Fateev {\it et al.} use the notation
$\lambda$ and $M$ which are related with the
parameters\ $\beta^2$\ and\ $ a$\  as
following, 
$$ \lambda=1-\beta^2\, , \ \ \ \ M=4\,  a^{-1}\ .$$}
The fields\ ${\bf T}^2$\ and ${\bar {\bf T}}^2$ have Lorenz spins
$4$ and $-4$ respectively. Therefore, the ``$\lambda_-$-term'' destroys
the rotational symmetry of the  Gaussian model in the infinite volume 
and appears because of the
lattice nature of the Hamiltonian\ \xxz.

\newsec{Temperature and magnetic field 
corrections to specific free energy}

The effective  Hamiltonian\ \sew\ can be applied  to the study
of infrared   corrections of physically interesting quantities in the
XXZ spin chain.
Here we deal with the  low temperature thermodynamics.
Let us consider the model\ \xxz\ at
the low  temperature\
$T\, \ (\, T/J\ll 1\, )$ and in the
weak magnetic field $h\, \  (\,  |h/J|\ll \, 1) $. 
With the result\ \sew\ one can develop the perturbation
expansion in the small parameter  $a$ of the 
specific free energy $f_{XXZ}$, 
\eqn\jshdgyt{e^{-{N\over T}\, f_{XXZ}}={\rm Tr}_{{\cal H}} \biggr[\,
e^{-{1\over T}\, {\bf H}_{XXZ}+{ h\over 2 T}\, {\bf S}^{z} }\, \biggl]\ . }
Notice  that ``$\lambda_{\pm}$-terms'' in \sew\ contribute
to the first order of the perturbation series and lead to
$a^3$-corrections.
The term with $\lambda$ does not have  diagonal matrix elements and
its contribution appears 
only in the second order. This implies
a correction which is  proportional to
$a^{{4\over \beta^2}-3}$.
Therefore  the leading corrections to the specific
free energy   
differ significantly in the domains\ 
$0<\beta^2<{2\over 3}$ and ${2\over 3}<\beta^2<1$ and have the form
\eqn\sjdhgy{\eqalign{
&f_{XXZ}={\cal E}_{0}+a\,  f_{Gauss}-
{a^3\over 2\pi}\ \Big( \lambda_+ 
\langle\langle\, {\bf T}\, \rangle\rangle^2+2\lambda_-
\langle\langle\, {\bf T}^2\, \rangle\rangle \Big)+...\ \ \ \ \ 
{\rm for}\ \ 0<\beta^2<{2\over 3}\ ,
\cr &
f_{XXZ}= {\cal E}_{0}+ a\, f_{Gauss}-\cr&
\ \ \ \ \ \ \ \ \ \ \     
{\lambda^2 a^{{4\over \beta^{2}}-3}\over 16\pi^2}\, 
\int_{0}^{L}dx
\int_{0}^{{1\over T}}d\tau \, \langle\langle\,
e^{i \tilde\varphi/\beta}(x,\tau)e^{-i \tilde \varphi/\beta}(0,0)
\, \rangle\rangle
+... \ \ \ \ \
{\rm for}\
{2\over 3}< \beta^2<1\ .}}
Here we use the notation
$$\eqalign{&e^{-{L\over T}\, f_{Gauss}}={\rm Tr}_{{\cal H}} \biggr[\,
e^{-{1 \over T}\, {\bf H}_{Gauss}+
{ h\over 2\beta T}\, (P-{\bar P}) }\, \biggl]\ ,\cr
&\langle\langle\, {\bf O}\, \rangle\rangle=
{\rm Tr}_{{\cal H}} \biggr[\,
e^{-{1 \over T}\, {\bf H}_{Gauss}+
{ h\over 2\beta T}\, (P-{\bar P}) }
\ {\bf O}\, \biggl]\Big/
{\rm Tr}_{{\cal H}} \biggr[\,
e^{-{1 \over T}\, {\bf H}_{Gauss}+
{ h\over 2\beta T}\, (P-{\bar P}) }
\, \biggl]\ ,}
$$
and
$$\tilde{\varphi}(x,\tau)=\phi(x+i\tau)+{\bar \phi}(x-i\tau)\ . $$

The calculation of\ \sjdhgy\ is straightforward.
We will not discuss here the 
most general expressions but restrict our attention to
the case $T\gg L^{-1}.$
Under this condition the system  develops  an effective
correlation length and all finite size effects are  suppressed by the
exponentially small factor $e^{-\pi{L T}}$. Omitting the finite size terms,
one  obtains
\eqn\ksjtaaa{f_{XXZ}={\cal E}_{0}+ 4\pi a\,T^2\
I_1\big({i h\over 8\pi\beta T}\big)- (2\pi a)^3 \,  T^4\
\Big\{ \lambda_+\, I^2_1
\big({i h\over 8\pi\beta T}\big)+
2\lambda_-\,  I_3
\big({i h\over 8\pi\beta T}\big) \Big\}+...\  .}
for $0<\beta^2<{2\over 3}$. If ${2\over 3}<\beta^2<1 $,
the leading correction  comes from the second order perturbation 
theory term
\eqn\ksjt{\eqalign{f_{XXZ}={\cal E}_{0}&+ 4\pi a\, T^2
\ 
I_1\big({i h\over 8\pi\beta T}\big)-\cr &
4 a\, T^2\,  \beta^8\sin\big({2\pi\over \beta^2}\big)\
\biggr[\,   {aT \sqrt\pi\,  \Gamma\big(1+{\beta^2\over
2-2\beta^2}\big)\over   \Gamma
\big(1+{1\over
2-2\beta^2}\big)}\, \biggl]^{{4\over\beta^{2}}-4}\
\Big| {\tilde H}_1\big({i h\
\over 8\pi\beta T}\Big)\Big|^2+...\ .}}
The auxiliary functions in\ \ksjtaaa,\ \ksjt\ read explicitly
\eqn\jsdgtt{\eqalign{
&I_1(p)=p^2-{1\over 24}\ , \cr 
&I_3(p)=p^4-{p^2\over 4}\, -
{4\beta^4-17\beta^2+4
\over 960 \beta^2}\ ,\cr
&{\tilde H}_1(p)={\Gamma\big(\beta^{-2}\big)\, 
\Gamma\big(1-2\beta^{-2}\big)\over\beta^4 \Gamma\big(1-\beta^{-2}\big)}\ 
{\Gamma\big( \beta^{-2}+2 p\beta^{-1}\big)\over
\Gamma\big(1- \beta^{-2}+2 p\beta^{-1}\big)\ }\ }}
and $\lambda_{\pm}=\lambda_{\pm}(\beta^2)$ are given by\ \jsdhgyt.
Notice that $I_1(p),\ I_3(p)$ coincide with
the vacuum eigenvalues of the first local integrals of motion in
the  quantum
KdV theory\ \BLZ. At the same time ${\tilde H}_1(p)$
is a vacuum  eigenvalue of the first  ``dual unlocal'' integral
of motion\ \BLZZ.

If\ $\beta^2={2 n\over 1+2 n}\ ( n=1,2,...)$,
the first order contribution of  the ``$\lambda_+$-term'' 
interferes with the $2 n$ order of the  perturbation
expansion generated by  the ``$\lambda$-term''.
The poles of the
coupling constant $\lambda_+$\ \jsdhgyt\ at these points  indicate
this phenomena. The interference
produces a logarithmic contribution. For  example, at $\beta^2={2\over 3}$
the specific free energy behaves as
\eqn\skdjdy{\eqalign{f_{XXZ}\big|_{\beta^2={2\over 3}}=
-{2\over 3\sqrt3\, a}&-{a\over 96 \pi}\ \big(16\pi^2  T^2 + 9 h^2 \big)
\ \biggr\{ 1+ {5\over 6}\, a^2T^2+ {a^2\over 192 \pi^2}\
\big( 16\pi^2 T^2+ 9 h^2 \big)\times\cr
&\biggl[\, {11\over 12}-\log\big(a T/12\big)-
\Re e\,\Big(
\psi\big({1\over 2}+{3 i h\over 8 \pi T}\big)\Big)\, \biggr]
\biggl\}+...\ ,   }}
where $\psi(t)=\partial_t\log\big(\Gamma(t)\big)$.

If $\beta^2\approx 1$,
Eq.\ksjt\ defines   
the leading asymptotic behavior  in the  very narrow domain of
temperature
and magnetic field,
$$ \log\big({1\over a T}\big)\, ,\ \    
\log\big({1\over a h}\big)\gg {1\over 1-\beta^2}\ .$$
In the  limit
$\beta^2\to 1$ the domain of  validity of \ksjt\ disappears completely.
Hence the case of the XXX spin chain requires
an additional analysis similar to the one of Refs.\ \sosage,\ 
\Zarn.

Let us consider the Gaussian model with $\beta^2=1$.
The  fields
\eqn\jshdt{\eqalign{&J_0={1\over 2}\,  \partial_x\phi\, , \ \ \ \
J_{\pm}=e^{\pm i\phi}\, ,\cr
&{\bar J}_0= {1\over 2}\, \partial_x{\bar \phi}\, , \ \ \ \
{\bar J}_{\pm}=e^{\mp i{\bar \phi}}\, , }}
generate the right and left current algebras at the  level $k=1$, and
the Gaussian model
coincides with
the  Wess-Zumino-Witten (WZW) model,
\eqn\hsgdr{{\bf H}_{Gauss}\Big|_{\beta^2=1}
={\bf H}_{WZW}\ .}
In the vicinity of the point $\beta^2=1$ it is  convenient
to rewrite the effective XXZ Hamiltonian as the marginal current-current
perturbation of the WZW Hamiltonian\ \afa,
\eqn\csdee{{\bf H}_{XXZ}=a^{-1} L\,  {\cal E}_0+ {\bf H}_{WZW}+
\int_{0}^{L} {d x\over 2\pi}\ \Big\{\, g_{\parallel}\, J_0 {\bar J}_0+
{g_{\perp}\over 2}\, \big( J_+ {\bar J}_-
+J_- {\bar J}_+\big)+...\,  \Big\}\, .}
Here $g_{\parallel}\geq    0$ and $|g_{\perp}|\leq g_{\parallel}$ are
small running coupling
constants. The corresponding Renormalization Group (RG) equations are
known exactly\ \Zarn,\ \zamol.
Under a suitable diffeomorphism of $g_{\parallel}$
and $g_{\perp}$ (i.e. the renormalization scheme choice),
one can set up
\eqn\jshdt{\eqalign{&r\,{dg_{\parallel}\over dr}=-
{2\, g_{\perp}^2\over 2-g_{\parallel}}\ ,\cr
&r\, {dg_{\perp} \over dr}=-
{2\, g_{\parallel} g_{\perp}\over 2-g_{\parallel}}\ .}}
Here  $r$ is the RG length scale.
These equations are solved as
\eqn\hsydt{g_{\parallel}=2\, (1-\beta^2) \ {1+q\over 1-q}\ ,\ \ \ \ \ 
g_{\perp}=-4\, (1-\beta^2) \ {q^{1\over 2}\over 1-q}\ ,}
with
\eqn\jsydr{ q\ (1-q)^{{2\over \beta^2}-2}=
\Big(\, {r\over r_0}\, \Big) ^{4-{4\over \beta^2}}\ ,}
while  $r_0$ is  a   $r$-independent
constant.

The correction to the
specific free energy, descended 
from the ``$\lambda$-term'' of\ \sew,\  admits   a  power series expansion,
\eqn\ksudyt{\eqalign{
f_{XXZ}={\cal E}_{0}&-{a\pi  T^2\over 6}\, 
\Big\{1+u_{10}\  g_{\parallel}+u_{20}\  g_{\parallel}^2+
u_{02}\ g_{\perp}^2+O(g^3) \Big\}-\cr &
{ah^2\over16\pi}\Big\{
1+v_{10}\  g_{\parallel}+v_{20}\ g_{\parallel}^2+
v_{02}\ g_{\perp}^2+O(g^3) \Big\}+...\ .}}
To find  values of the first coefficients
$u_{kj}, v_{kj}$ we should   compare
the expansion of\ \ksudyt\ and\ \ksjt\ in
power series of $1-\beta^2$ for a 
fixed value of the parameter $q$. According to\ \jsydr,
the latter   can be chosen as a
solution of the algebraic equation
\eqn\skdjyt{ q\ (1-q)^{{2\over \beta^2}-2}=
\biggr[\,   {aT\, e^{-{1\over 4}} \sqrt\pi\,  \Gamma\big(1+{\beta^2\over
2-2\beta^2}\big)\over   \Gamma
\big(1+{1\over
2-2\beta^2}\big)}\, \biggl]^{{4\over\beta^{2}}-4}\ \biggl|
{\Gamma\big(\beta^{-2}+ {ih\over 4\pi\beta^2 T}\big)\over
\Gamma\big(2-\beta^{-2}+ {i h\over 4\pi\beta^2 T}\big)}
\biggr|^2\ . }
A simple calculation leads to the result
\eqn\hsgdtt{\eqalign{
f_{XXZ}={\cal E}_{0}&-{\pi a T^2 \over 6}\, \Big\{
1+{3\over 8}\,  g_{\parallel}g_{\perp}^2+O(g^4)\Big\}-\cr &
{ah^2\over16\pi}\Big\{
1+{g_{\parallel}\over 2}+
{g_{\parallel}^2\over 4}-{g_{\perp}^2\over 4}+
{g_{\parallel}^3\over 8}-{g_{\parallel}g_{\perp}^2\over 32}+
O(g^4)\Big\}+...\ .}} 
Now we can take the limit\ $\beta^2\to 1$, 
\eqn\vseu{ g_{\parallel}\to g\ ,\ \ \ \ g_{\perp}\to- g\  ,}
of  Eqs.\hsydt,\  
\skdjyt,\ \hsgdtt\ and obtain
the low temperature behavior of the  specific free energy of  the
XXX spin chain,
\eqn\ksjdh{\eqalign{f_{XXX}=&-2J \, \log(2) -
{T^2\over 6 J}\,
\Big\{ 1+{3\over 8}\, g^3+O(g^4) \Big\}-
{h^2\over 16 \pi^2  J}\, \Big\{  1+{g\over 2}+{3\over 32}\, g^3+
O(g^4) \Big\}-\cr&
{\sqrt3\over 16 \pi^3 J^3}\ \Big\{\,  {h^4\over 64\pi^2} +
{h^2T^2\over 4} +  
{3\pi^2 T^4\over 5}+ O(g)\, \Big\}+...\ .}}
The function  $g=g(T,h)$ in\ \ksjdh\ solves
the equation
\eqn\jsudt{g^{-1}+{1\over 2}\, \log(g)=-\Re e\Big(\psi\big(1+
{i h\over 4\pi T}\big)\Big)+\log
\big(\sqrt{2\pi} e^{1\over 4} J/T\big)\ ,}
with $\psi(t)=\partial_t\log\big(\Gamma(t)\big)$.
In the formulas\ \ksjdh,\ \jsudt\ we use the original
parameter $J$ of the Hamiltonian\ \xxz\ with $\Delta=1$.
We also include the correction arising from
the ``$\lambda_-$-term''\ \ksjtaaa.
{}From\ \jsudt\  one has the formula for
the magnetic
susceptibility,
\eqn\hsgdr{\chi=-4\ \partial^2_h f_{XXX}\big|_{h=0}\ ,}
as a function of temperature,
\eqn\shdgt{\chi(T)={1\over 2 J  \pi^2} \Big\{1+{g\over 2}+
{3\, g^3\over 32}+O(g^4)+
{\sqrt3\, T^2\over 4 \pi  J^2}\, \big(1+O(g)\big)+...\, \Big\}\ .}
Now the running constant depends  on temperature only,\ $g=g(T)$,
$$g^{-1}+{1\over 2}\, \log(g)=
\log
\big(\sqrt{2\pi} e^{\gamma+{ 1\over 4}} J/T\big)\  $$
and $\gamma=0.577216...$ is the Euler constant.
The leading  $T$-dependence (the first order in $g$)
of  $\chi(T)$ was obtained previously in\ \Af.
In Fig.$\,1$, \shdgt\ is compared against the result of
numerical solution of
the Thermodynamic Bethe Ansatz equations\ \Af.
\midinsert

\centerline{\epsfbox{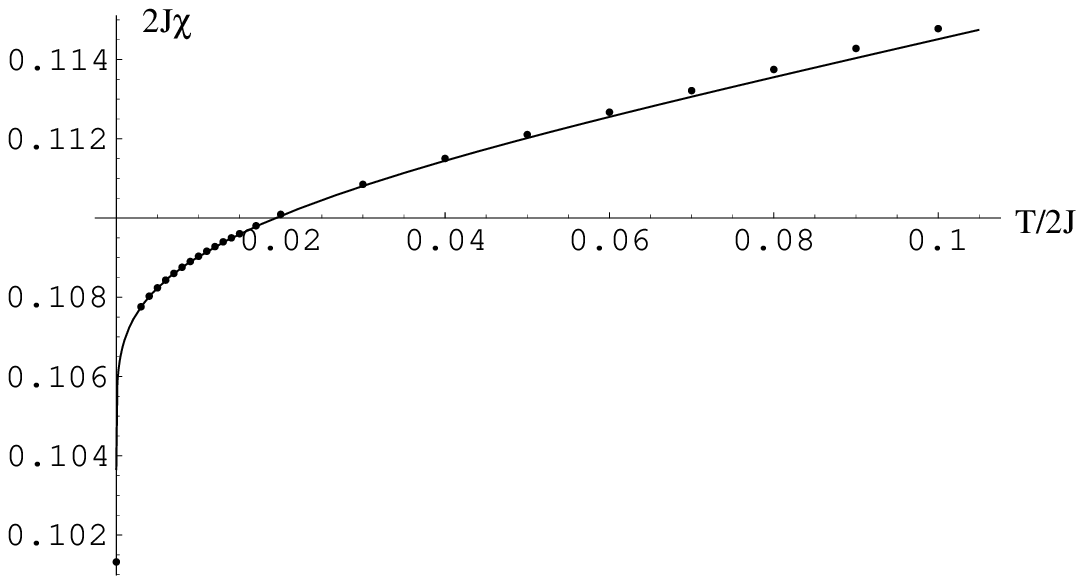}}
\vskip 7 truemm
\capt{
Fig.$\,1$. Magnetic susceptibility of the XXX antiferromagnetic spin chain.
The continuous line represents\ \shdgt. The bullets were obtained by
the numerical solution of the Thermodynamic Bethe Ansatz equations\ \Af. }
\vskip 7 truemm
\endinsert

\newsec{Finite size corrections to vacuum energies}

In this Section we consider the XXZ spin chain at zero magnetic
field and temperature.
If $T=0$, the function $f_{XXZ}$\ \jshdgyt\ coincides with the specific
ground state
energy $E_{XXZ}^{(0)} /N$ and\ Eqs.\sjdhgy\ are useful  
to study finite size effects. In fact,  the calculation
can be easily
generalized  to obtain the finite size corrections 
to an arbitrary vacuum eigenvalue (not only ground state)
$E_{XXZ}^{(s)}$ in the
sector with a given value $s$ of the total spin\ ${\bf S}^z$\ \usyt.
Let us parameterize $E_{XXZ}^{(s)}/N$ by the form,
\eqn\hsgdtr{
E^{(s)}_{XXZ}/N=  {\cal E}_{0}+
{2\pi  \over a N^2 }\  \delta(s,\theta,N)\ , }
Then,
\eqn\hdt{\delta={s^2\beta^2\over 2}+
{\theta^2\over 2\beta^2}-{1\over 12}+
\delta^{{\cos}}+\delta^{{\bf T}}+...\ . }
The function\ $\delta_{{\bf T}}$ gives
the  first order contribution of the
``$\lambda_{\pm}$-terms'' in\ \sew.
It reads explicitly,\foot{
In writing\ \hdt, we assume that $\beta^2\not={2 n\over 2n+1}\, ,
\ n=2,3,...$ \ (see comment to the  equation\ \skdjdy).} 
\eqn\shdtr{\delta^{{\bf T}}=-
{4 \pi^2\over N^2}
\Big\{
\lambda_+\,  I_1\big({s\beta\over 2}+{\theta\over 2\beta}\big)\,
I_1\big({s\beta\over 2}-{\theta\over 2\beta}\big)+
\lambda_-
\, \Big(\, I_3\big({s\beta\over 2}+{\theta\over 2\beta}\big)+
I_3\big({s\beta\over 2}-{\theta\over 2\beta}\big)\,\Big)\Big\}\ .}
The term $\delta^{{\cos}}$ in\ \hdt\  arises from
the second order perturbation theory generated by the
field\ $\cos({\tilde \varphi}/\beta)$,
\eqn\gsfrru{\delta^{{\cos}}=-
{2\over \pi}\, \beta^8\, \sin\big({2\pi\over \beta^2}\big)\,
\biggl[\,{\sqrt\pi   \Gamma(1+{\beta^2\over 2-2\beta^2})
\over  N   \Gamma
(1+{1\over
2-2\beta^2}}\, \biggl]^{{4\over\beta^{2}}-4}\
{\tilde H}_1\big({s\beta\over 2}+{\theta\over 2\beta}\big)\,
{\tilde H}_1\big({s\beta\over 2}-{\theta\over 2\beta}\big)\ .}
The auxiliary  functions  in\ \shdtr,\ \gsfrru\ are given by \jsdgtt.
For $\beta^2$ sufficiently close to unity the finite size corrections
require an additional RG consideration.
In a similar way to\ \hsgdtt, an appropriate RG analysis leads to the
following expression for\ \hdt,
\eqn\gdfr{\delta(s,\theta,N)=\delta^{RG}+\delta^{{\bf T}}+...\ .}
The function $\delta^{RG}$ is
a  RG improved contribution of the
``$\lambda$-term'',
\eqn\jsydt{\eqalign{
\delta^{RG}=&-{1\over 12}\ \Big\{1+{3\over 8}
\,  g_{\parallel}g_{\perp}^2\Big\}+
{s^2\over 2}\ \Big\{
1-{g_{\parallel}\over 2}+{1\over 4}\, g_{\perp}^2-
{7\over 32}\, g_{\parallel}g_{\perp}^2
\Big\}+
{|s|\over 16} \ \big\{ 2 g_{\perp}^2-g_{\parallel}g_{\perp}^2
\big\}+\cr &
{\theta^2\over 2}\ \Big\{
1+{g_{\parallel}\over 2}+
{g_{\parallel}^2\over 4}-{g_{\perp}^2\over 4}+
{g_{\parallel}^3\over 8}-
{g_{\parallel}g_{\perp}^2\over 32}\Big\}+
O(g^4)\ , }}
where the running constants are given by\ 
\hsydt\ with 
\eqn\hsgdfr{\eqalign{ q\ & (1-q)^{{2\over \beta^2}-2}=
\biggr[\,   {\sqrt\pi\,e^{-{1\over 4}}  \Gamma\big(1+{\beta^2\over
2-2\beta^2}\big)\over N\,   \Gamma
\big(1+{1\over
2-2\beta^2}\big)}\, \biggl]^{{4\over\beta^{2}}-4}\times\cr & 
\ \ {\Gamma\big(\beta^{-2}+ s+\theta\beta^{-2}\big)
\, \Gamma\big(\beta^{-2}+ s-\theta\beta^{-2}\big)\over
\Gamma\big(2-\beta^{-2}+ s+\theta\beta^{-2}\big)
\, \Gamma\big(2-\beta^{-2}+ s-\theta\beta^{-2}\big) }\ .}}
The function
$$\delta(s,\theta,N)-\delta(0,0,N)\ $$
was  calculated in\ \barber\ by the numerical solution of the Algebraic
Bethe Ansatz equations for $\beta^2={5\over 6}$ and
some vacuum energies.
In Table 1  the numerical  results
are compared against the derived formula.
The following comment is appropriate here. The function\ \gsfrru, as
well as the right hand side of\ \hsgdfr, has the pole at  $s=0,\ \theta=1$. 
The reason of the singularity is  rather simple; 
We have developed the perturbation theory  starting with the
Fock vacuum states\ $|\,p,{\bar p} \,
\rangle$\ \hsgdreww, with
$$ p={\theta\over 2\beta}+{s\over 2\beta}\ , \ \ \ 
{\bar p}={\theta\over 2\beta}-{s\over 2\beta}\ .$$
But for $s=0$ and
$\theta$   sufficiently close to unity,  the contribution 
of the vacuum state  with $p={\bar p}={\theta-2\over 2\beta}$  to
the true  ground state of the XXZ spin chain  becomes significant.
It seems reasonable  that
the disagreement  between the numerical data and
obtained asymptotics for the case $s=0, \ \theta={1\over 2}$ 
in   Table 1 is due to this effect.

The finite size corrections
for the XXX spin chain can be derived by taking the limit\ \vseu\
of  Eqs.\jsydt, \shdtr. In particular,
\eqn\jshdgtt{\eqalign{&\delta^{{\bf T}} \big|_{\beta^2=1}=
-{\pi\sqrt3\over 8 N^2}\ \Big\{
s^4+\theta^4+6\, s^2 \theta^2-s^2-\theta^2+
{3\over 20}\Big\} \ , \cr
&\delta^{RG} \big|_{\beta^2=1}=-{1\over 12}\ \Big\{1+{3\over 8} \,
 g^3\Big\}+
{s^2\over 2}\ \Big\{1-{g\over 2}+{g^2\over 4}-{7\over 32}\,
g^3\Big\}
+{|s|\over 16}\  \big\{ 2 g^2-g^3\big\}+\cr &
\ \ \ \ \ \ \ \ \ \ \
{\theta^2\over 2}\ \Big\{1+{g\over 2}+{3 g^3\over 32}\Big\}+
O(g^4)\ ,}}
with
$$g^{-1}+{1\over 2}\, \log(g)=\log
\big(2^{{1\over 2}}\pi^{-{1\over 2}} e^{1\over 4} N\big)-
{1\over 2}\Big( \psi\big(1+s+\theta)+
\psi\big(1+s-\theta) \Big)\ . $$
The leading terms
in\ \jshdgtt\  for\ $\theta=0,\  s=0,1$\  are
in agreement with the results of the work\ \AGSZ.
\midinsert
\bigskip
\centerline{
\noindent\vbox{\offinterlineskip
\def\tablerule{\noalign{\hrule}}
\halign{ 
\strut#&\vrule#\tabskip=1em plus2em& 
   #&\vrule\vrule#&
   #&\vrule#&
   #&\vrule\vrule#& 
   #&\vrule#&
   #&\vrule\vrule#&
   #&\vrule#& 
   #&\vrule#
\tabskip=0pt
\cr\tablerule
&& 
&&&\omit\hidewidth $s=0,\, \theta={1\over 2}$\hidewidth &&  
&&\omit\hidewidth  $s=1,\, \theta=0$\hidewidth &&
&&\omit\hidewidth  $s=1,\, \theta={1\over 4}$\hidewidth &&
\cr\tablerule
&& $N$  && \ \ BA  && \ RG &&\ \   BA &&\  RG &&\ \   BA && \ RG &
\cr\tablerule
&&8 && 0.15519  && 0.15467 &&  0.39793 && 0.395... && 0.43606 && 0.433... &
\cr\tablerule
&&16 && 0.15217  && 0.15185 &&  0.40594 && 0.40555 && 0.44460 && 0.44406 &
\cr\tablerule
&&32 && 0.15104  && 0.15081 &&  0.41063 && 0.41060 && 0.44897 && 0.44883 &
\cr\tablerule
&&64 && 0.15055  &&  0.15037 &&  0.41327 && 0.41327 && 0.45128 && 0.45121 &
\cr\tablerule
&&128 && 0.15030 && 0.15015 &&  0.41474 &&  0.41474 && 0.45254 && 0.45250 &
\cr\tablerule
&&256 && 0.15017  && 0.15003 &&  0.41557 && 0.41557 && 0.45324 && 0.45321 &
\cr\tablerule
&&$\infty$
&&&\omit\hidewidth 0.15000 \hidewidth &&
&&\omit\hidewidth  0.41600 \hidewidth &&
&&\omit\hidewidth  0.45416 \hidewidth &&
\cr\tablerule
\noalign{\smallskip} }
}
}
\bigskip
\capt{Table 1. 
The function $\delta(s,\theta,N)-\delta(0,0,N)$ for $\beta^2={5\over 6}$. 
The ``BA''
columns were obtained  by means of the  numerical solution of the 
Algebraic Bethe Ansatz equations\ \barber.
The ``RG'' columns follow from\ \gdfr,\ \jsydt,\ \shdtr.  
}
\endinsert

\newsec{Subleading asymptotics of correlation functions}

As well as the  Hamiltonian density\ \sew,\ the formulas\ \shdttt\ 
give only leading terms of 
the expansions in  local quantum fields of the
spin operators.
The  leading terms allow one to obtain the leading asymptotic
behavior of  correlation functions of the spin operators.
Subleading asymptotics are the result of the two effects:
next terms in the expansion\ \shdttt\ and
XXZ corrections of the Gaussian ground state, which  defined by
the effective Hamiltonian\ \sew.
Therefore, to derive  the subleading
asymptotics of the  correlation functions systematically,  one
needs to know  next terms
in the expansion\ \shdttt. Unfortunately, this problem admits  only 
a qualitative analysis at the moment.
Nevertheless, some interesting 
quantitative predictions  for the  subleading
asymptotics
can be obtained based  on
the  effective
Hamiltonian\ \sew\ only. As an example 
we consider the  equal time   correlator
\eqn\kshjdgy{W(n)={{\langle\, vac\, |\,
\sigma_{k}^{x}\, \sigma_{k+n}^{x}\, |\, vac\,
\rangle }\over{\langle\, vac\, |\, vac \, \rangle }}\, , \ \ \ \ \ \ 
n=1,2...\ ,}
for the infinite spin chain $(L=\infty)$.
Simple  arguments\ \afa\ show that
the most important correction of the  bosonization formula for
$\sigma_{k+n}^{\pm}$ has the form
\eqn\hsdttra{\sigma_{k+n}^{\pm}=a^{{\beta^2\over 2}}\
\sqrt{{F\over 2}}\
e^{\pm i{\beta\varphi\over 2} }(x)\ \Big(1\pm  i\, (-1)^n \ a\ A\, 
\partial_x \varphi(x)+...
\Big)\  \ \ \ (\, x=n a\, )\ . }
The exact value of the  coefficient $A$ is not known nowadays.
It is instructive   to note  that  
due to the sign factor $(-1)^n$ in \hsdttra\ a 
linear contribution
of the ``$A$-term'' in $W(n)$  is canceled out exactly
for odd  integer $n$ and
the subleading asymptotic is determined by
the effective Hamiltonian only.
We conclude  that for ${2\over 3}<\beta^2<1$,
\eqn\hsgdt{W(n)=F\ n^{-\beta^2}\ \Big\{\, 1- C\
n^{4-{4\over\beta^2}}-2\beta A\, \big( (-1)^n+1\big)\ n^{-1}+...\, 
\Big\}\ ,  \ \ \ \
{\rm as}\ n\to+\infty\ .}
Here the constant  $F$ is given by\ \xxzfactor,
$A=A(\beta)$ is not known,
while the constant $C=C(\beta^2)$
can be obtained by the
second order perturbation theory based on the effective Hamiltonian\ \sew.
The calculation leads to the formula
\eqn\shdtre{\eqalign{C=& 
{ \Gamma^2(\beta^{-2})\over
\Gamma^2 (1-\beta^{-2})}\ \biggr[\,{ \Gamma\big(1+
{\beta^2\over 2-2\beta^2}\big)
\over 2\, \sqrt\pi \ \Gamma\big(1+{1\over 2-2\beta^2}\big)}\, \biggl]^{
{4\over \beta^{2}}-4}\  
\Big(\, {2\, \pi^2\over \sin^2\big(2\pi\beta^{-2}\big)}-\cr & 
{ \beta^4\over (1-\beta^2)(2-\beta^2)}-\psi'\big(\beta^{-2}\big)-
\psi'\big({3\over 2}-\beta^{-2}\big)\, \Big)\ ,}}
with\  $\psi'(t)=\partial^2_t\log\big(\Gamma(t)\big)$.
Eq.\hsgdt\ defines the subleading 
asymptotic behavior of the correlation function
for
$$ \log(n)\gg {1\over 1-\beta^2}\ .$$
In order to study the limit $\beta^2\to 1$, it is useful to 
rearrange  the expansion of $W(n)$.
To do it, let us introduce $g_{\parallel}$ and $ g _{ \perp}$ 
by Eqs.\hsydt, while
\eqn\hsgdr{q\  (1-q)^{{2\over \beta^2}-2}=
\biggr[\,   {e^{-\gamma-1}\,  \Gamma\big(1+{\beta^2\over
2-2\beta^2}\big)\over 2\,\sqrt\pi\, n\,   \Gamma
\big(1+{1\over
2-2\beta^2}\big)}\, \biggl]^{{4\over\beta^{2}}-4}\ ,}
and $\gamma$ is the Euler constant.
Using\ \hsgdt,\ \shdtre, one  obtains
\eqn\xssss{\eqalign{
W(n)=  \sqrt{{2 \over \pi^3}}\ 
{Z(g_{\parallel}, g _{ \perp})\over n\, \sqrt{|\, g_{\perp}|  } }\ 
& \Big\{1-{g_{\parallel}^2\over 8}-
{g _{ \perp}^2\over 16}
-{4\, \zeta(3)+3\over 192}\ g_{\parallel}^3+
{164\, \zeta(3)-67
\over 384}\  g_{\parallel} g^2_{\perp}
+O(g^4)-\cr &
\  {{\cal A}\over 2 n}\ 
\big(\, (-1)^n+1+O(g)\, \big)+O(n^{-2})\, \Big\}\ ,}}
where\ ${\cal A}=4\beta\, A\big|_{\beta^2=1}$ and  $\zeta(s)$ is 
the Riemann zeta function.
The non-perturbative function $Z$ in\ \xssss\ reads
$$Z(g_{\parallel}, g _{ \perp})
=\biggr[{|\, g_{\perp}|\over 2\,e^{1\over 2}\, g_{\parallel}}
\biggl]^{\sqrt{g^2_{\parallel}-g^2_{\perp}}\big/4 }\ .$$
We can derive the  asymptotic of the correlation function of the
XXX spin chain by taking the limit
\ $g_{\parallel}\to g\, ,\ \  g_{\perp}\to -g\ .$
Finally one has\foot{
The coefficient $\sqrt{{2 \over \pi^3}}$ in
Eq.(5.7)\ was derived previously  in\ \affleck\  based on
the explicit form of the constant $F$\ \xxzfactor. I am grateful to
I. Affleck for communications concerning this calculation.}  
\eqn\yugdr{\eqalign{ W(n)\big|_{\beta^2=1}=
\sqrt{{2 \over \pi^3}}\  {1\over n\,  \sqrt{g}}\ & 
\Big\{1-
{3\over 16}\, g^2+ {156\, \zeta(3)-73\over 384}
\ g^3+O(g^4)-\cr &
\ { {\cal A} \over 2 n}\
\big(\, (-1)^n+1+O(g)\, \big)+O(n^{-2})\, \Big\}\ ,}}
and
$$g^{-1}+{1\over 2}\, \log(g)=\log
\big(2\sqrt{2\pi} e^{\gamma+1} n \big)\ .$$

The correlation function\ $W(n)\big|_{\beta^2=1}\  (1\leq n\leq 30)$\ was
calculated numerically in the works\ \lin,\ \hallberg.\foot{
K.A. Hallberg {\it et al.}\ \hallberg\ used the notations
$\omega(n)=W(n)/4\big|_{\beta^2=1}$.}
In  Table 2 the numerical data are compared against\ \yugdr\ 
for odd integers $n$. 
Unfortunately,  we can not directly compare the numerics
for even integers $n$, since the  value of
the constant ${\cal A}$ in\ \yugdr\
is not known. Fitting  of the  data gives
\eqn\jsudt{{\cal A}\approx 0.4\ .}  
The corresponding plots are  presented on Fig. 2.
\midinsert

\centerline{\epsfbox{ 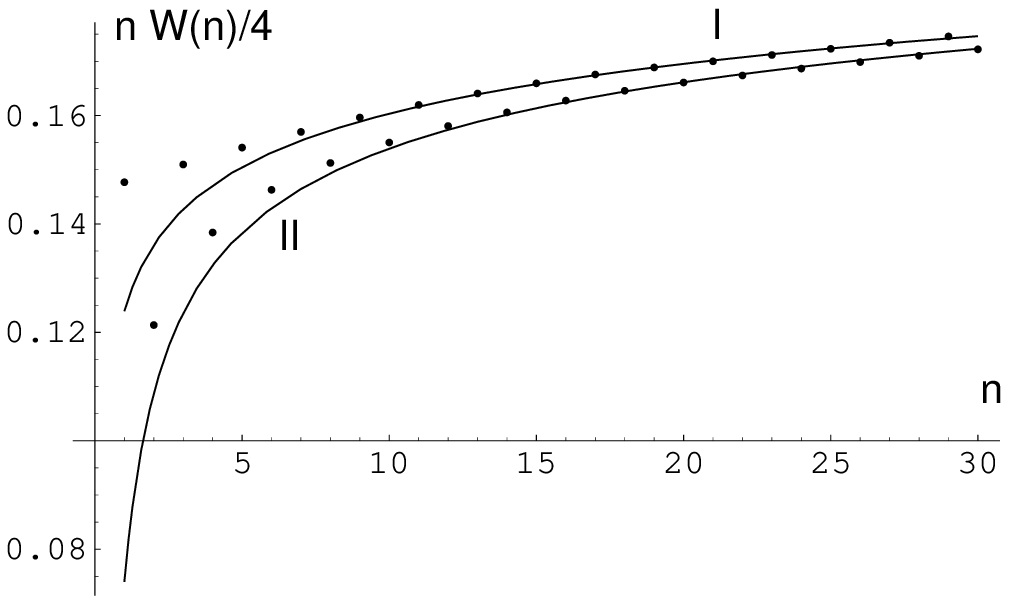 }}
\vskip 5 truemm
\capt{
Fig.$\,2$. The correlation  function\ ${n\over 4}\
{{\langle\, vac\, |\,
\sigma_{k}^{x}\, \sigma_{k+n}^{x}\, |\, vac\,
\rangle }\over{ \langle\, vac\, |\, vac \, \rangle }} $\
for the XXX spin chain.
The bullets
(see Table 2) were obtained in\ \hallberg.
The continuous line $I$ follows from\  
\yugdr\ for odd integers $n$. The  line $II$ represents\ \yugdr\ for
even\ $n$\ ($\, {\cal A}=0.4\, $). }
\vskip 5 truemm
\endinsert

Finally we note that there are two different approaches
to exact calculation of
correlation functions like\ \kshjdgy\ at the moment.
In the  first one
the correlation functions  are expressed in terms of
Fredholm determinants\ \Kor,\ \Korr,\ \Korra.
Recently  M. Jimbo and T. Miwa\ \MiW\ found
another representation for  the correlation functions in terms of
$n$-fold integrals. It would be interesting to derive
the large\ $n$\ asymptotics\ \hsgdt,\ \yugdr\ via
these approaches.

\centerline{}

\centerline{\bf Acknowledgments}

\vskip0.5cm

I am grateful to  I. Affleck, Al.B. Zamolodchikov,
A.B. Zamolodchikov for sharing their insights and
K.A. Hallberg who kindly provided  me with  
the numerical data which was not
included in the paper\ \hallberg.
The work was started during the visit at SdPT ${\grave {\rm a}}$ Saclay.
I am grateful
to the members of the laboratory and   especially D. Bernard and
J.-B. Zuber for their kind hospitality and interesting discussions.
I thank  D. Fradkin and  A. Morosov for many explanations on
computer software.
The research  is supported in part by the DOE grant \#DE-FG05-90ER40559.

\vfill
\eject

\midinsert
\centerline{
\noindent\vbox{\offinterlineskip
\def\tablerule{\noalign{\hrule}}
\halign{
\strut#&\vrule#\tabskip=1em plus2em&
   #&\vrule#&
   #&\vrule#&
   #&\vrule#
\tabskip=0pt
\cr\tablerule
&& $n$  && \ \ NUM  && \ RG &
\cr\tablerule
&& 1  && 0.1477  && 0.1239 &
\cr\tablerule
&& 2  && 0.1214  &&  &
\cr\tablerule
&& 3  && 0.1510  && 0.1427 &
\cr\tablerule
&& 4  && 0.1384  &&  &
\cr\tablerule
&& 5  && 0.1541  && 0.1505  &
\cr\tablerule
&& 6  && 0.1463  &&  &
\cr\tablerule
&& 7  &&  0.1567  && 0.1554  &
\cr\tablerule
&& 8  &&  0.1513   &&  &
\cr\tablerule
&& 9  && 0.1596  && 0.1589 &
\cr\tablerule
&& 10  && 0.1550   &&  &
\cr\tablerule
&& 11  && 0.1620  && 0.1616  &
\cr\tablerule
&& 12  && 0.1581   &&  &
\cr\tablerule
&& 13  && 0.1641  && 0.1639 &
\cr\tablerule
&& 14  && 0.1606  &&  &
\cr\tablerule
&& 15  && 0.1659  && 0.1658 &
\cr\tablerule
&& 16  && 0.1628  &&  &
\cr\tablerule
&& 17  && 0.1676  && 0.1674 &
\cr\tablerule
&& 18  &&  0.1646  &&  &
\cr\tablerule
&& 19  && 0.1689  && 0.1689 &
\cr\tablerule
&& 20  && 0.1661  &&  &
\cr\tablerule
&& 21  && 0.1700  && 0.1702 &
\cr\tablerule
&& 22  && 0.1674  &&  &
\cr\tablerule
&& 23  && 0.1712 && 0.1713 &
\cr\tablerule
&& 24  && 0.1687  &&  &
\cr\tablerule
&& 25  && 0.1723  && 0.1724 &
\cr\tablerule
&& 26  && 0.1699  &&  &
\cr\tablerule
&& 27  && 0.1734  && 0.1733 &
\cr\tablerule
&& 28  && 0.1710  &&  &
\cr\tablerule
&& 29  &&  0.1746  && 0.1742 &
\cr\tablerule
&& 30  && 0.1722  &&  &
\cr\tablerule
\noalign{\smallskip} }
}
}
\bigskip
\capt{Table 2. The correlation function ${n\over 4}\
{{\langle\, vac\, |\,
\sigma_{k}^{x}\, \sigma_{k+n}^{x}\, |\, vac\,
\rangle }\over{ \langle\, vac\, |\, vac \, \rangle }} $\
for the XXX spin chain. 
The column ``NUM'' were obtained in\ \hallberg.
The column ``RG'' follows from\ \kshjdgy\ and\  \yugdr\ for 
odd integers $n$.}
\endinsert
\vfill
\eject

\newsec{Appendix}

The calculation of  the  coupling
constants in the effective Hamiltonian\ \sew\ is based on the
technique developed in the works\ \Hasenf,\ \sosage, \Zarn.
We discuss it very briefly.

Let us consider the XXZ spin 
chain\ \xxz\ with infinite number of sites $N$
in the   magnetic field $h$\ \jshdgyt. 
The specific ground  state energy 
\eqn\ksjhdt{{\cal E}(h)=\lim_{N\to \infty}\, E^{(0)}_{XXZ}(h)/N\, , }
reads\ \yangaa,\ \Kor
\eqn\jsdgt{ {\cal E}(h)=-{h\over 4}+J\, 
\int_{-\Lambda}^{\Lambda}{d\alpha\over 2\pi}
\ {\sin(\pi\beta^2)\  \epsilon(\alpha)\over \cosh(2\alpha)+
\cos(\pi\beta^2)}\ ,}
where $\epsilon(\alpha)$ satisfies  the Yang-Yang equation
\eqn\jsudt{\eqalign{\epsilon(\alpha)-\int_{-\Lambda}^{\Lambda}{
d\alpha'\over \pi}\ &
{\sin(2\pi\beta^2)\ \epsilon(\alpha')\over 
\cosh(2\alpha-2\alpha')-\cos(2\pi\beta^2)}\ 
={h\over J}-{4\sin^2(\pi\beta^2)\over 
\cosh(\alpha)+\cos(2\pi\beta^2)}\ ,\cr 
&\ \ \ \ \ \ \ \ \ \ \ \ \ \ \ \epsilon(\pm \Lambda)=0\ .}}
Solving\ \jsudt\ by the  Wiener-Hopf method,
we obtain for the 
weak magnetic field
\eqn\sjduyy{{\cal E}(h)={\cal E}_0-{ a\, h^2
\over 16\pi\beta^2}\ 
\big(\, 1+\kappa\ \big|\, a h\, \big|^{{4\over \beta^2}-4}+
(\kappa_+ +\kappa_-)\  (\, a h\, )^2+...\, \big)\ \ \ \ 
(\,   |h/J|\ll 1  )\ ,}
where the constant $a$ and ${\cal E}_0$ are
given by\ \hsgdta,\ \jsdhfgr, and
\eqn\shdgrtr{\eqalign{
&\kappa={\Gamma^2(\beta^{-2})\, 
\tan\big(\pi\beta^{-2}\big)\over 2\ \beta^2\,
\Gamma^2({1\over 2}+\beta^{-2})}\ 
\biggr[\,{ \Gamma\big(
{\beta^2\over 2-2\beta^2}\big)
\over 8\, 
\sqrt\pi \ \Gamma\big({1\over 2-2\beta^2}\big)}\, \biggl]^{
{2\over \beta^{2}}-
2}\ ,\cr
&\kappa_+={1\over 64 \pi \beta^2 }\, 
\tan\big({\pi\over 2-2\beta^2}\big)\ ,\cr
&\kappa_-={1\over 192\pi}\ {\Gamma\big({3\over 2-2\beta^2}\big)\,
\Gamma^3\big({\beta^2\over 2-2\beta^2}\big) \over
\Gamma\big({3\beta^2\over 2-2\beta^2}\big)\,
\Gamma^3\big({1\over 2-2\beta^2}\big)}\  . }}
Now, we should  compare\ \sjduyy\  with the result of 
the perturbation theory based on the effective Hamiltonian\ \sew.
This allows to  extract the values of the constant
$\lambda$ and the  combination $\lambda_+ +2\lambda_-$.
The  ``$\lambda_+$'' and ``$\lambda_-$-terms'' in
the effective Hamiltonian\ \sew\
differ essentially. The operator\
${\bf T}{\bar {\bf T}}$\  has Lorenz spin $0$, while
${\bf T}^2\ ({\bar {\bf T}}^2)$ has Lorenz spin $4 \, (-4)$.
Hence, the  $\lambda_+$-term preserves 
the Lorenz invariance,\ whereas
the ``$\lambda_-$-term''
destroys
the rotational symmetry.
A careful analysis of the
expansion\ \sjduyy\ based on this  simple observation 
makes it  possible to split  contributions of the operators\
${\bf T}{\bar {\bf T}}$\ and ${\bf T}^2+ {\bar {\bf T}}^2$\ and
leads to\ \jsdhgyt.

\listrefs

\end